\ifdefined\pdfminorversion\pdfminorversion=7\fi
\documentclass[sigconf,9pt,nonacm]{acmart}

\setcopyright{none}
\hypersetup{hidelinks}

\usepackage{graphicx}
\usepackage{amsmath}
\usepackage{booktabs}
\usepackage{xspace}
\usepackage{multirow}
\usepackage{xcolor}
\usepackage{algorithm}
\usepackage{algpseudocode}
\usepackage{makecell}
\graphicspath{{figures/}}
\title[Building Fully On-Device GUI Agents with Mobile NPUs]{Building Fully On-Device GUI Agents \texorpdfstring{\\}{ }with Mobile NPUs}
\author{Zhe Dong}
\affiliation{%
  \institution{Tsinghua University}
  \city{Beijing}
  \country{China}}
\email{dong-z22@mails.tsinghua.edu.cn}

\author{Guanyu Cai}
\affiliation{%
  \institution{Tsinghua University}
  \city{Beijing}
  \country{China}}
\email{cgy22@mails.tsinghua.edu.cn}

\author{Yuchen Jiang}
\affiliation{%
  \institution{Beijing Institute of Technology}
  \city{Beijing}
  \country{China}}
\email{jiangycs@bit.edu.cn}

\author{Jiliang Wang}
\affiliation{%
  \institution{Tsinghua University}
  \city{Beijing}
  \country{China}}
\email{jiliangwang@tsinghua.edu.cn}

\begin{document}
\begin{abstract}
Mobile GUI agents are increasingly used to automate everyday smartphone tasks, but most existing systems still rely on cloud-hosted models or remote inference, introducing repeated communication latency, privacy exposure, and dependence on network availability. Modern smartphones integrate neural processing units (NPUs) that provide an attractive substrate for fully on-device LLM inference. However, directly deploying conventional GUI-agent pipelines on mobile NPUs is inefficient. GUI-agent inference is fundamentally a long-input, short-output workload, yet conventional agents spend substantial time on autoregressive reasoning and action generation, while mobile NPUs provide much higher throughput for prompt prefill than token-by-token decoding. Meanwhile, dynamically varying GUI-agent contexts conflict with the fixed context profiles required by statically compiled NPU runtimes.
We present \emph{DroidPrefill}, a prefill-centric inference architecture
that reshapes the GUI-agent workload around mobile-NPU execution.
DroidPrefill combines \emph{Prefill-Centric Inference} to reduce
unnecessary autoregressive decoding with \emph{Efficient Context
Management} to preserve decision-critical information within a fixed
compiled context budget. Experiments on commercial smartphones show that,
in controlled comparisons with the autoregressive T3A+ baseline, DroidPrefill
reduces per-step inference latency by up to 85.0\%, while improving
task success rate by up to 24.6\%.

\end{abstract}

\maketitle
\hypersetup{pdftitle={Building Fully On-Device GUI Agents with Mobile NPUs},
  pdfauthor={Zhe Dong, Guanyu Cai, Yuchen Jiang, Jiliang Wang}}
\pagestyle{plain}
\begin{figure}[t]
    \centering
    \includegraphics[width=\linewidth]{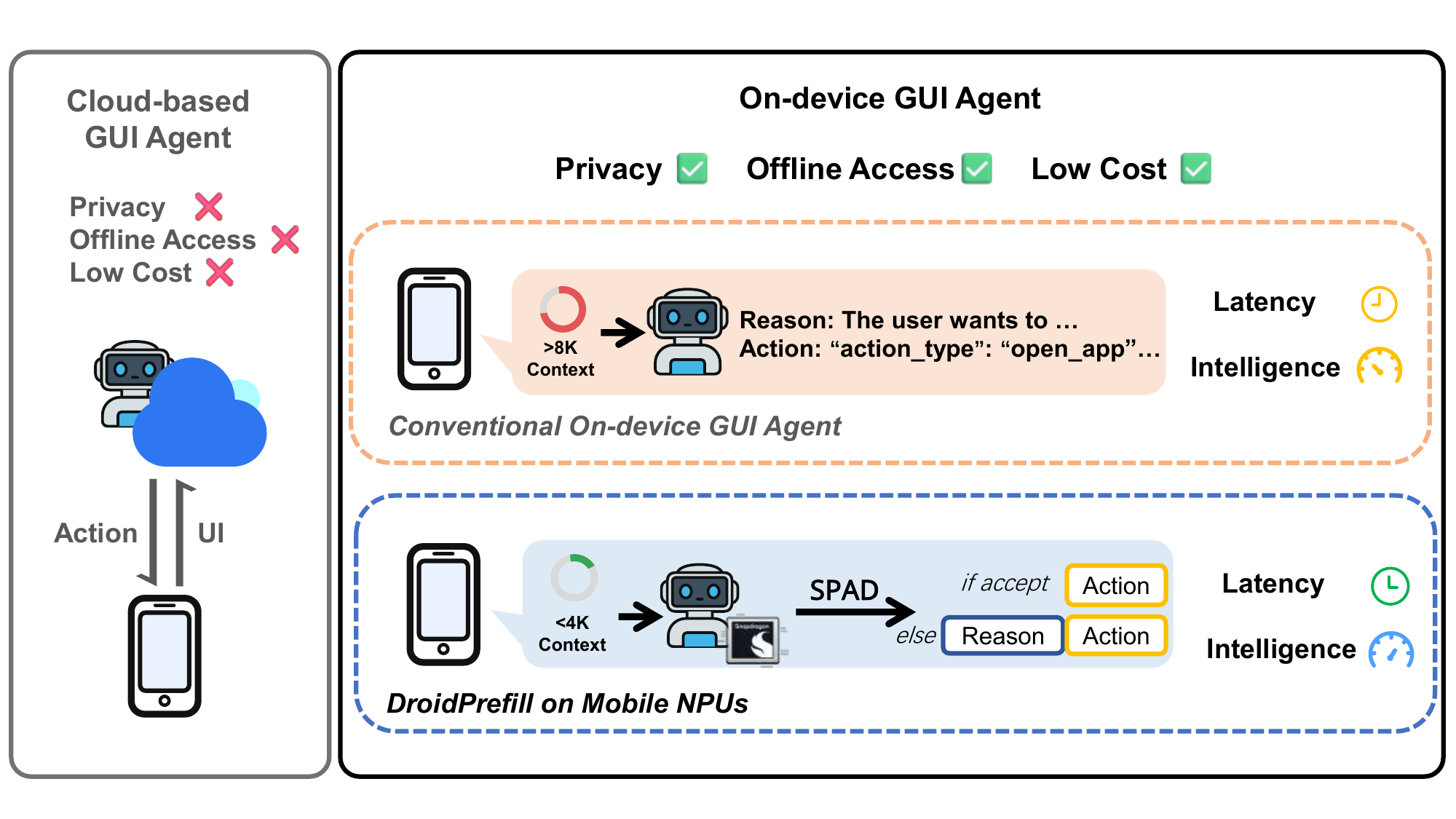}
    \caption{Comparison of cloud-based, conventional on-device, and
    DroidPrefill GUI agents. Moving inference on device improves privacy,
    offline availability, and cost, but a conventional on-device agent
    still inherits the long-context, decode-heavy inference pipeline of
    cloud agents, resulting in high latency on mobile devices. DroidPrefill
    instead reshapes the workload through efficient context management and
    prefill-centric inference, reducing latency while preserving agent
    capability.}
    \label{fig:overview}
\end{figure}

\section{Introduction}
\label{sec:introduction}
Mobile GUI agents provide a general-purpose approach to automating
everyday smartphone tasks, such as navigating applications, retrieving
information, and completing forms, while also improving application
accessibility~\cite{autoglm2024,autodroid2024,mobilegpt2024,appagent2025,
androidworld2025}.
Unlike application-specific scripts, a GUI agent observes the current
interface, reasons about task progress, and executes user-level actions
such as tapping, typing, and scrolling.
Recent advances in large language models (LLMs) and
vision-language models (VLMs) have enabled increasingly capable
mobile GUI agents~\cite{uitars2025,agentcpmgui2025,
stepgui2025,phonebuddy2026}.
However, many existing systems still rely on cloud-hosted models
or remote inference.
Because multi-step tasks repeatedly invoke an
observe--reason--execute loop, cloud-based agents incur network
communication and remote-inference latency at every step, expose
privacy-sensitive UI states, and remain dependent on network
availability~\cite{autodroidv2_2025,mobileexplorer2026}.
These limitations motivate fully on-device mobile GUI
agents~\cite{yuan2024firmware,agentcpmgui2025,stepgui2025}.

Modern smartphones integrate heterogeneous compute resources, including
CPUs, GPUs, and neural processing units (NPUs) for local
machine-learning workloads~\cite{jeong2022band,yuan2024firmware}.
Among them, NPUs are particularly attractive for mobile agents because
they provide specialized and energy-efficient acceleration for LLM
inference while preserving CPU and GPU resources for application logic
and rendering~\cite{llmnpu2025,shadownpu2026,npuready2026}.

Figure~\ref{fig:overview} contrasts cloud-based GUI agents, a
conventional on-device deployment, and DroidPrefill. Although moving
inference on device improves privacy, offline availability, and cost,
a conventional on-device agent still inherits the long-context,
decode-heavy inference pattern of cloud agents, making it inefficient
on mobile NPUs.

We observe that GUI-agent inference is fundamentally a
\emph{long-input, short-output} workload.
Each interaction step may consume a long task instruction, UI
representation, and interaction history, yet typically produces only
one executable action from a small action space~\cite{androidworld2025}.
This structure creates a natural opportunity for mobile NPUs: the long
input can be processed during high-throughput prefill, while only a
small amount of autoregressive decoding is fundamentally required.

Guided by this observation, we present \emph{DroidPrefill}, a
prefill-centric inference architecture for fully on-device mobile
agents on commercial smartphone NPUs.
Rather than asking the model to freely generate a reasoning trace and
action at every interaction step, \emph{DroidPrefill} first compiles the
current UI state into a compact set of executable candidates.
\emph{Efficient Context Management} then preserves UI
evidence and bounded task state within a fixed NPU context budget.
After this context is prefilled, \emph{Prefill-Centric Inference} directly
examines the candidate-label logits exposed at the prefill boundary.
Confident decisions can be dispatched immediately through
\emph{Self-Speculative Action Dispatch} (SPAD), while uncertain decisions
continue from the same KV cache through a short Structured Decision
Refinement step.
Only actions that require textual arguments invoke an additional
bounded fill stage.

Together, these mechanisms shift common-case decision making away from
long autoregressive generation while keeping dynamically varying agent
contexts within an efficient compiled profile.
They directly address two workload--hardware mismatches that prevent
conventional mobile-agent pipelines from efficiently exploiting mobile
NPUs:

\textbf{\textit{Prefill--decode mismatch.}}
Mobile NPUs execute prompt prefill efficiently, whereas autoregressive
decoding is sequential and substantially slower
~\cite{llmnpu2025,npuready2026}.
Conventional GUI agents, however, spend substantial inference time
decoding reasoning traces and free-form actions.
DroidPrefill addresses this mismatch by moving common-case action
selection to the prefill boundary and invoking decoding only when
additional reasoning or textual output is needed.

\textbf{\textit{Static--dynamic mismatch.}}
Mobile NPU runtimes commonly rely on statically compiled context
profiles, while GUI-agent request lengths vary dynamically across tasks,
UI states, and interaction steps.
Using a large profile reduces inference efficiency, whereas a small
profile may fail to accommodate context-heavy requests.
DroidPrefill addresses this mismatch by packing decision-critical UI,
task-state, and action information within a bounded context budget.

In summary, this paper makes the following contributions:

\begin{itemize}
    \item To the best of our knowledge, we present the first systematic
    study of mobile GUI-agent inference on commercial smartphone NPUs.
    We identify two critical workload--hardware mismatches: a
    \emph{prefill--decode mismatch} between decode-heavy conventional
    agents and prefill-efficient NPU execution, and a
    \emph{static--dynamic mismatch} between dynamically varying
    GUI-agent contexts and statically compiled context profiles.

    \item We design DroidPrefill, an NPU-native mobile GUI-agent
    architecture that combines \emph{Prefill-Centric Inference} with
    \emph{Efficient Context Management}. DroidPrefill moves common-case
    action selection to the prefill boundary and packs decision-critical
    context within an efficient compiled context profile.

    \item We implement DroidPrefill on commercial smartphone NPUs and
    evaluate its task effectiveness and inference efficiency.
    Across the evaluated deployments, DroidPrefill reduces per-step
    inference latency by up to 85.0\% while improving task
    success rate by up to 24.6\%.
    
\end{itemize}

\section{Motivation}
\label{sec:background}

\subsection{GUI Agent Inference Workload}
\label{sec:workload}

A mobile GUI agent repeatedly maps a user task, the current UI state,
and interaction history to the next executable action.
Unlike a single-turn language-model request, this mapping is invoked
repeatedly throughout a task: each executed action changes the
application state, producing a new observation and therefore a new
inference request.
We model one interaction step as
\begin{equation}
    a_t = \pi_{\theta}(q, s_t, h_t),
    \label{eq:agent-step}
\end{equation}
where $q$ is the user task, $s_t$ is the current UI observation,
$h_t$ is the interaction history or compact task state, and $a_t$ is
the next executable GUI action.
The observation $s_t$ may contain an accessibility hierarchy, UI XML,
visible text, actionable-element metadata, and an optional screenshot.
A task is completed through a closed loop of observation, inference,
execution, and state update.

These inputs exhibit different degrees of reuse and variation across
interaction steps. The task instruction $q$ typically remains unchanged,
whereas the UI observation $s_t$ changes after each executed action and
the interaction state $h_t$ evolves with task progress. Consequently,
each model invocation repeatedly combines a relatively stable task
context with a potentially large, step-specific GUI state. In contrast,
the required output is usually compact: the agent only needs to determine
the next executable operation and, when required, an associated textual
argument.

A conventional GUI agent request typically combines the task
instruction, UI observation, accumulated interaction history, and
output-format instructions.
The model then autoregressively generates a reasoning trace followed by
an action.
This cloud-oriented inference pattern is poorly matched to mobile
deployment: GUI observations are long, tasks require repeated
interaction steps, and free-form reasoning introduces substantial
autoregressive generation.

Modern mobile NPUs provide specialized and energy-efficient acceleration
for neural workloads while leaving CPU and GPU resources available to
foreground applications.
However, our measurements show that conventional GUI-agent inference
does not naturally match mobile-NPU execution.
We identify two key mismatches: a large cost gap between prompt prefill
and autoregressive decode, and a conflict between dynamically varying
GUI-agent requests and statically provisioned context capacity.

\subsection{Prefill--Decode Mismatch}
\label{sec:prefill-decode}

Transformer inference consists of two phases.
During \emph{prefill}, the model processes the prompt and constructs the
key--value (KV) cache.
Because all prompt tokens are available in advance, prefill exposes
substantial parallelism.
During \emph{decode}, the model generates one token at a time, with each
token depending on the previously generated sequence.
Prefill and decode therefore exhibit fundamentally different execution
characteristics on mobile accelerators.

\begin{figure}[t]
    \centering
    \includegraphics[width=\linewidth]{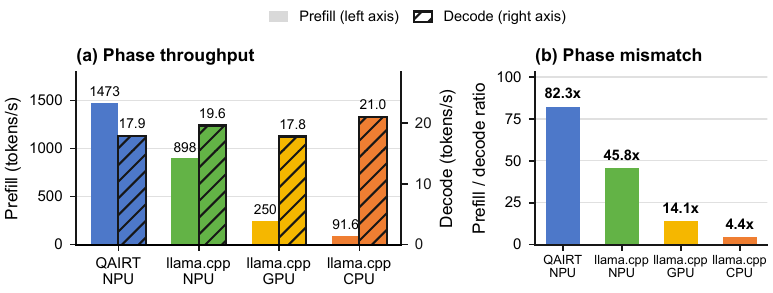}
    \caption{Prefill--decode measurements for
    Qwen3-4B-Instruct-2507. Solid bars denote parallel prompt prefill and
    hatched bars denote autoregressive decode. NPU paths expose the
    largest phase mismatch, while phone CPU/GPU provide much lower
    prefill throughput.}
    \label{fig:prefill-decode}
\end{figure}

Figure~\ref{fig:prefill-decode} shows this mismatch across mobile
inference runtimes.
At the 128-token operating point, QAIRT/W4A16 reaches 1,473 prompt
tokens/s but only 17.9 decoded tokens/s, an $82.3\times$ ratio.
llama.cpp/Q4\_0 reaches 897.7 and 19.6 tokens/s,
respectively, still a $45.8\times$ ratio.
Although the runtime changes the absolute operating point, the
qualitative result remains the same: processing additional information
during prefill is substantially cheaper than generating additional
tokens autoregressively.

An additional observation is that the NPU advantage is concentrated
primarily in the prefill phase. Across the evaluated backends, decode
throughput remains within a relatively narrow range, whereas prefill
throughput differs substantially. This means that simply moving a
decode-heavy agent from CPU/GPU execution to the NPU does not fully
exploit the accelerator. The agent pipeline must instead expose more
parallelizable work to prefill and reduce the amount of sequential
generation on the critical path.

This asymmetry is particularly important for GUI agents.
A GUI step may consume a rich state description, including the UI,
task state, and interaction history, but ultimately needs to produce
only one executable action.
A conventional CoT agent instead generates tens or hundreds of
reasoning tokens before emitting that action.
Moreover, recent empirical evidence shows that reasoning-enabled models
provide only marginal and inconsistent improvements for mobile GUI
agents and can even degrade performance under some configurations
~\cite{zhang2026cotgui}.
Thus, unconditional CoT introduces a substantial inference cost without
a guaranteed improvement in task effectiveness.

Across a task with $T$ interaction steps, the inference latency can be
expressed conceptually as
\begin{equation}
    T_{\mathrm{model}}
    \approx
    \sum_{t=1}^{T}
    \left(
        T_{\mathrm{prefill},t}
        +
        N_{\mathrm{decode},t}
        \cdot
        t_{\mathrm{decode}}
    \right),
    \label{eq:task-latency}
\end{equation}
where $N_{\mathrm{decode},t}$ is the number of generated tokens at
step $t$, and $t_{\mathrm{decode}}$ denotes the average per-token decode
latency.

To interpret this asymmetry, let $P_{\mathrm{prefill}}$ and
$P_{\mathrm{decode}}$ denote the prefill and decode throughput,
respectively, and define
\begin{equation}
    \rho =
    \frac{P_{\mathrm{prefill}}}
         {P_{\mathrm{decode}}}.
    \label{eq:prefill-decode-ratio}
\end{equation}
Under a constant-throughput approximation, one decode token incurs
approximately the same latency as prefilling $\rho$ additional input
tokens.
With the QAIRT backend on the smartphone NPU,
$\rho \approx 82.3$, so eliminating one decode token can offset
the latency of roughly 82 additional prefill tokens.
This suggests that minimizing total token count alone is not the right
objective on mobile NPUs; the agent should instead reduce the
phase-weighted inference cost induced by the prefill--decode mismatch.

Together, these observations motivate moving common-case GUI action
selection toward the prefill boundary and reserving autoregressive
decoding for cases that truly require additional reasoning or textual
generation.

\begin{figure}[t]
    \centering
    \includegraphics[width=\linewidth]
    {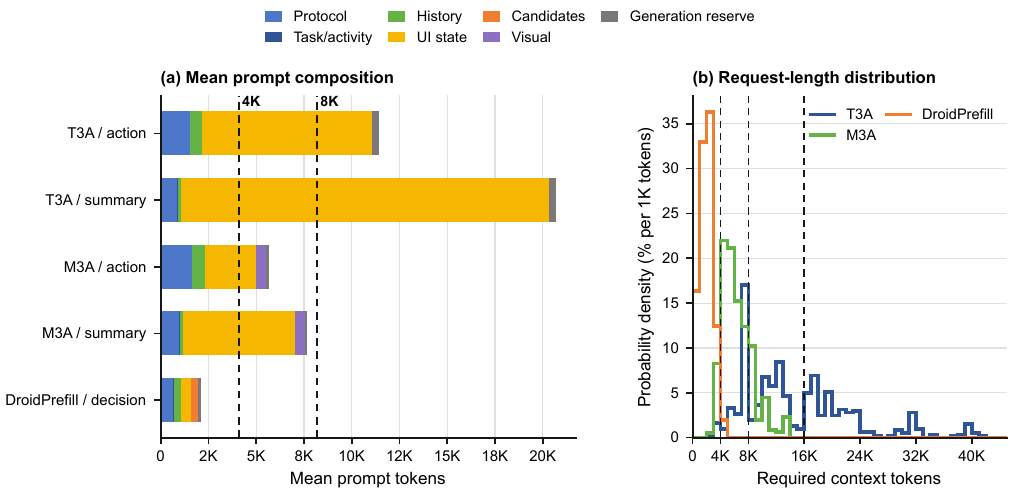}
    \caption{Prompt study over AndroidWorld traces.
    Panel~(a) decomposes mean prompt length; panel~(b) shows the
    normalized empirical density of the complete request stream using
    uniform 1K-token bins. Dashed vertical lines mark 4K, 8K, and 16K
    context profiles.}
    \label{fig:prompt-breakdown}
\end{figure}

\subsection{Static--Dynamic Mismatch}
\label{sec:fixed-profile}

The second mismatch arises because GUI agent request lengths vary at
run time, whereas mobile-NPU deployment commonly fixes the maximum
context capacity in advance.

We refer to such a deployment configuration as a
\emph{context profile}. A context profile is provisioned for a fixed
maximum context capacity $C_{\max}$ at deployment time, while the
actual request length $C_{\mathrm{req},t}$ can vary across interaction
steps as long as it remains within that capacity. For example, we
evaluate 4K, 8K, and 16K context profiles, corresponding to different
values of $C_{\max}$.

Formally,
\begin{equation}
    C_{\mathrm{req},t} \leq C_{\max},
    \quad
    C_{\max}\ \text{fixed},
    \quad
    C_{\mathrm{req},t}\ \text{variable}.
    \label{eq:motivation-static-dynamic}
\end{equation}

Unlike runtime-configured CPU/GPU backends, mobile-NPU deployment
commonly relies on ahead-of-time compiled or otherwise provisioned
context profiles~\cite{llmnpu2025,npuready2026}.
The deployment must therefore choose $C_{\max}$ before observing the
step-specific UI state and history that determine
$C_{\mathrm{req},t}$.

To quantify this mismatch, we reconstruct matched prompts from the same
AndroidWorld traces, covering 361 decision states and 244 inter-step
summary calls.
Every request is measured using the same Qwen tokenizer and its
configured generation reserve.
We evaluate canonical T3A and M3A requests together with DroidPrefill
decision requests.

Figure~\ref{fig:prompt-breakdown} shows that conventional GUI-agent
requests are both long and highly variable.
Raw T3A action and summary requests average 11.4K and 20.7K tokens;
only 1.8\%, 25.8\%, and 59.3\% of the complete request stream fit 4K,
8K, and 16K profiles.
Raw M3A action and summary requests average 5.7K and 7.6K tokens,
yielding 8.8\%, 79.5\%, and 100.0\% coverage.
Thus, a small context profile rejects or truncates many conventional
requests, while a large profile is required to cover their long tail.

\begin{figure}[t]
    \centering

    \includegraphics[width=\linewidth]{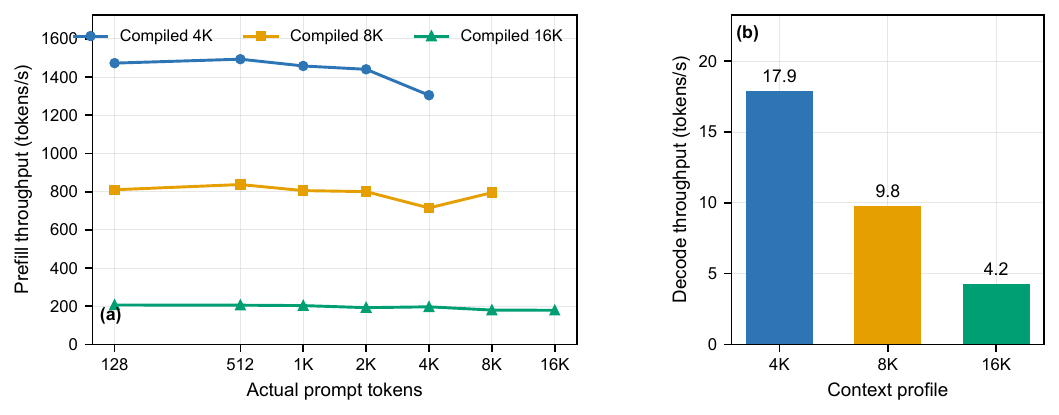}
    \vspace{2pt}

    \includegraphics[width=\linewidth]{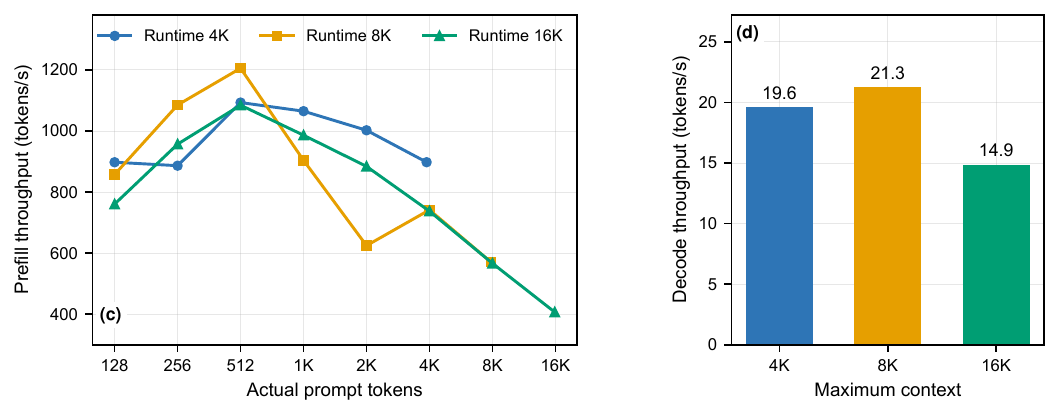}

    \caption{Context scaling under different mobile inference runtimes.
    Top: QAIRT/W4A16 compiled context profiles. Bottom:
    llama.cpp/Q4\_0 runtime-configured context limits.
    In each row, the left panel reports prefill throughput versus actual
    prompt length, while the right panel reports decode throughput for
    128-token prompts.}
    \label{fig:context-scaling}
\end{figure}

A larger $C_{\max}$, however, is not free.
Changing the compiled context capacity may alter KV-cache provisioning,
static tensor shapes, buffer allocation, memory pressure, and graph
scheduling.
Figure~\ref{fig:context-scaling} shows the resulting performance
tradeoff.
The 8K and 16K QAIRT/W4A16 configurations provide greater context
capacity but lower prefill and decode throughput even when
$C_{\mathrm{req}}$ is held constant.
The runtime-configured llama.cpp/Q4\_0 path shows a weaker
context-capacity penalty, although throughput still decreases as the
occupied KV cache approaches capacity.

These results expose a fundamental deployment tradeoff:
$C_{\max}$ must be large enough to accommodate dynamic GUI-agent
requests, but unnecessarily large profiles reduce NPU efficiency.
An efficient mobile GUI agent should therefore preserve
decision-critical information within a small, fixed context budget
rather than relying on an oversized deployment profile.

\subsection{Design Implications}
\label{sec:design-implications}

The preceding measurements lead to two design principles for NPU-native
mobile GUI agents.

\textbf{Prefill-Friendly.}
The agent should expose a compact action space whose semantic decisions
can be extracted at the prefill boundary.
Rather than generating actions in an open-ended format, executable
actions should be represented as grounded candidates with compact
labels, allowing prefill logits to provide an inexpensive signal of the
next action.

\textbf{Context within Budget.}
The task instruction, UI evidence, interaction state, candidate actions,
and output space should be explicitly managed under the target compiled
context profile.
The agent should preserve decision-critical information within a small
fixed context budget instead of relying on a larger and slower profile.

Guided by these principles, we design DroidPrefill, which combines
Prefill-Centric Inference with Efficient Context Management for
fully on-device GUI-agent execution on mobile NPUs.


\section{DroidPrefill Design}
\label{sec:design}

\begin{figure}[t]
    \centering
    \includegraphics[width=\linewidth]{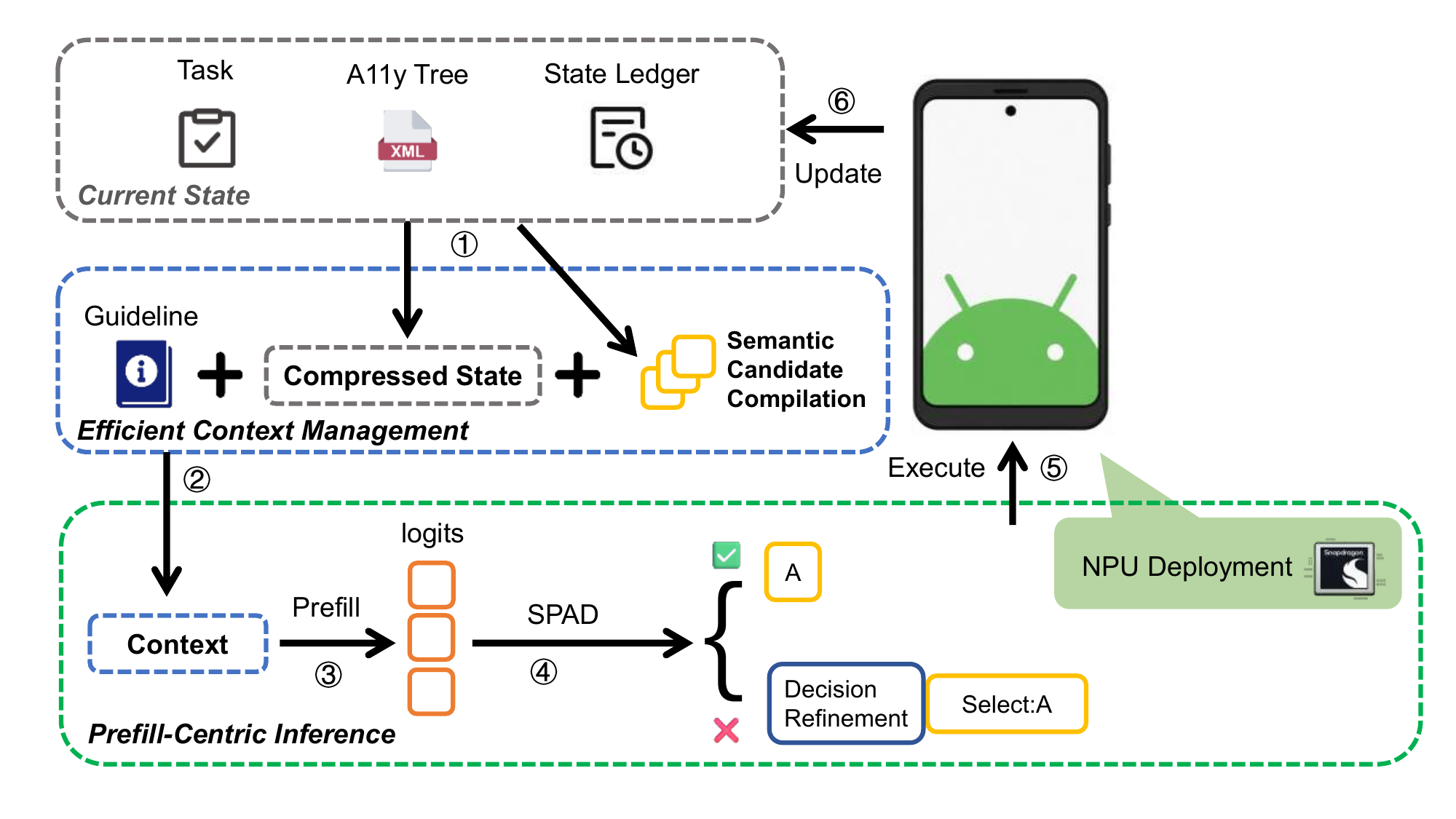}
    \caption{DroidPrefill architecture. At each interaction step,
    Semantic Candidate Compilation converts the current GUI state into
    grounded executable candidates. Efficient Context Management packs
    candidate-relevant UI evidence and bounded task state within the
    compiled NPU context budget. After prefill, Prefill-Centric
    Inference uses Self-Speculative Action Dispatch (SPAD) to execute
    confident decisions early, while uncertain decisions enter a short
    Structured Decision Refinement step using the same KV cache.
    Select-then-Fill generates a textual value only when required by the
    selected action.}
    \label{fig:architecture}
\end{figure}

\subsection{System Overview}
\label{sec:overview}

DroidPrefill is a fully on-device mobile GUI agent designed around the
long-input, short-output workload of GUI-agent inference.
Rather than asking the model to freely generate an operation, target,
argument, and long reasoning trace at every step, DroidPrefill first
converts the current GUI state into a finite set of grounded executable
candidates and constructs a bounded request for mobile-NPU inference.

At each interaction step, \emph{Semantic Candidate Compilation}
constructs executable candidates from the current UI state and binds
their targets to internal accessibility-node identifiers.
\emph{Efficient Context Management} then combines the task,
candidate-relevant UI evidence, bounded task state, and candidate
descriptions within the compiled context budget.
After the NPU prefills this request, \emph{Prefill-Centric Inference}
uses candidate-label logits to determine the next action.
Confident decisions can be dispatched through
\emph{Self-Speculative Action Dispatch} (SPAD), while uncertain
decisions enter a short \emph{Structured Decision Refinement} step using
the same KV cache.
If the selected action requires a textual value, DroidPrefill generates
only that value through \emph{Select-then-Fill}.

These components directly address the two mismatches identified in
Section~\ref{sec:background}.
Efficient Context Management keeps dynamically varying GUI-agent
requests within a fixed context profile, while Prefill-Centric
Inference reduces dependence on expensive autoregressive decoding.

Candidate construction, context construction, state management, model
inference, and action execution all run locally on the device.
No remote model or auxiliary summarization service participates in the
observation--decision--execution loop.

\subsection{Semantic Candidate Compilation}
\label{sec:candidate-compilation}

Conventional GUI agents expose an open-ended action space in which the
model must jointly generate the operation type, target, grounding, and
optional argument.
Prior mobile GUI agents have also used discrete or compact action spaces
to simplify grounding and action generation
~\cite{androidworld2025,vdroid2026}.
DroidPrefill builds on this idea by using grounded candidates as a shared
interface for both context construction and prefill-centric inference.
DroidPrefill then transforms the current GUI state into a finite set of
grounded executable candidates before model inference.

Candidate compilation is expressed as
\begin{equation}
    \mathcal{C}_t =
    F\!\left(
        q,\,
        L_t,\,
        G(s_t)
    \right),
    \label{eq:candidate-compilation}
\end{equation}
where $G(s_t)$ extracts the operations supported by the current UI
state, and $F$ applies task- and state-dependent filtering based on the
task instruction $q$ and the task-state ledger $L_t$.

Each candidate is represented as
\begin{equation}
    c_i =
    \left(
        k_i,\,
        g_i,\,
        d_i,\,
        \sigma_i
    \right),
    \label{eq:candidate-representation}
\end{equation}
where $k_i$ is the operation kind, $g_i$ is the internal grounding,
$d_i$ is the model-facing semantic description, and $\sigma_i$ is an
optional value slot.
The model never generates $g_i$ directly; the executor maintains this
binding internally.

\paragraph{Candidate generation and filtering.}
The compiler generates operations only for visible and enabled elements
that support the corresponding interaction.
Clickable, long-clickable, editable, and scrollable elements produce
the appropriate grounded operations.
Passive text is incorporated into the description of its actionable
ancestor, and candidates referring to the same executable target and
operation are deduplicated.

Candidate descriptions retain state information when it changes the
effect of an action.
For example, a stateful control may be represented as
\texttt{A. Tap [Wi-Fi switch; currently on; tapping turns it off]}
rather than simply \texttt{Tap [Wi-Fi]}.
The compiler also includes applicable global and terminal operations,
such as \textsc{OpenApp}, \textsc{Back}, \textsc{Home}, \textsc{Wait},
directional \textsc{Swipe}, \textsc{Complete}, and \textsc{Speak}.
Foreground activity and ledger state suppress redundant or stale
operations, such as reopening the current application or repeatedly
executing an action that produced no UI change.

\paragraph{Parameterized operations and compact labels.}
Parameterized operations retain an unresolved value slot.
For example, \textsc{Type} specifies an editable target but leaves its
text value unfilled, while \textsc{Speak} represents a terminal answer
whose content is generated only after selection.
After filtering, candidates receive compact labels from a
tokenizer-verified vocabulary:
\begin{verbatim}
A. Tap [Add alarm]
B. Tap [7:30 alarm; currently disabled]
C. Press Back
D. Complete the task
\end{verbatim}
Each label is represented by one token under the deployed tokenizer and
maps to exactly one internal executor payload for the current
interaction step.
Candidate semantics are therefore processed during prefill, while
action selection operates over compact labels.

\subsection{Efficient Context Management}
\label{sec:context-construction}

Given a compiled context capacity, DroidPrefill aims to preserve
decision-critical information rather than simply minimizing prompt
length.
Instead of assembling a complete prompt and truncating it afterward,
the system explicitly allocates the available budget across task
instructions, UI evidence, task state, candidate descriptions, and
structured decision output before serialization.

Let $C_{\max}$ denote the compiled context capacity.
At step $t$, DroidPrefill constructs request $x_t$ such that
\begin{equation}
    T_{\theta}(x_t)
    + R_{\mathrm{dec}}
    + R_{\mathrm{tpl}}
    \le C_{\max},
    \label{eq:context-budget}
\end{equation}
where $T_{\theta}(\cdot)$ is the deployed tokenizer,
$R_{\mathrm{dec}}$ reserves capacity for Structured Decision Refinement,
and $R_{\mathrm{tpl}}$ accounts for chat-template and runtime overhead.
A parameterized action, if selected, uses a separately bounded fill
request.

\subsubsection{Candidate-grounded UI compression.}

Direct serialization of an accessibility hierarchy contains substantial
information that is not needed for the next action decision, including
empty containers, duplicated descendants, absolute coordinates, and
framework-specific metadata.
Given the current candidate set $\mathcal{C}_t$, DroidPrefill instead
preserves UI evidence required to distinguish the candidate actions.

Nodes referenced by candidate targets are treated as mandatory evidence.
The compressor retains executable elements, their decision-relevant
states, and nearby labels needed to distinguish competing operations.
Passive labels are attached to their nearest actionable ancestor rather
than exposed as independent targets.
Element identifiers used by the executor are preserved internally so
that candidate descriptions and executable actions remain consistently
grounded.
Redundant structure and attributes unrelated to action selection are
removed.

\subsubsection{Bounded state ledger.}

A complete interaction transcript grows continuously with the task.
Growing interaction histories have also been identified as a key
context-management challenge for long-horizon mobile GUI agents
~\cite{agentprog2026}.
T3A-style agents commonly invoke the language model to summarize
previous actions and observations, introducing an additional inference
request after each step and allowing summary errors to propagate into
subsequent decisions.

DroidPrefill instead maintains a bounded ledger $L_t$ and updates it
deterministically from the observed transition
$(s_{t-1},a_{t-1},s_t)$.
The ledger records the current activity, recently executed operations,
observed UI changes, reached subgoals, and repeated ineffective actions.
Redundant or stale entries are merged or removed under a fixed entry
budget.
The update runs entirely in the local agent runtime and requires no
additional model invocation.

\paragraph{Cache-friendly assembly.}
DroidPrefill serializes task-invariant content before step-dependent
content:
\begin{equation}
    x_t =
    \left[
        x^{\mathrm{static}}(q)
        \,;\,
        x^{\mathrm{dynamic}}_t
    \right].
    \label{eq:static-dynamic-context}
\end{equation}
The static prefix contains the system instruction, action protocol,
output contract, task instruction, and fixed demonstrations.
The dynamic suffix contains the foreground activity, compressed UI,
state ledger, candidates, and selection query.
Deterministic formatting keeps the static portion byte-identical across
interaction steps, allowing its KV state to be reused when supported by
the runtime. Prompt restructuring for improved prefix-cache reuse has also been
explored for general on-device agent workloads~\cite{agentx2026}.

\subsection{Prefill-Centric Inference}
\label{sec:action-selection}
\label{sec:select_then_fill}
\begin{figure}[t]
    \centering
    \includegraphics[width=\linewidth]{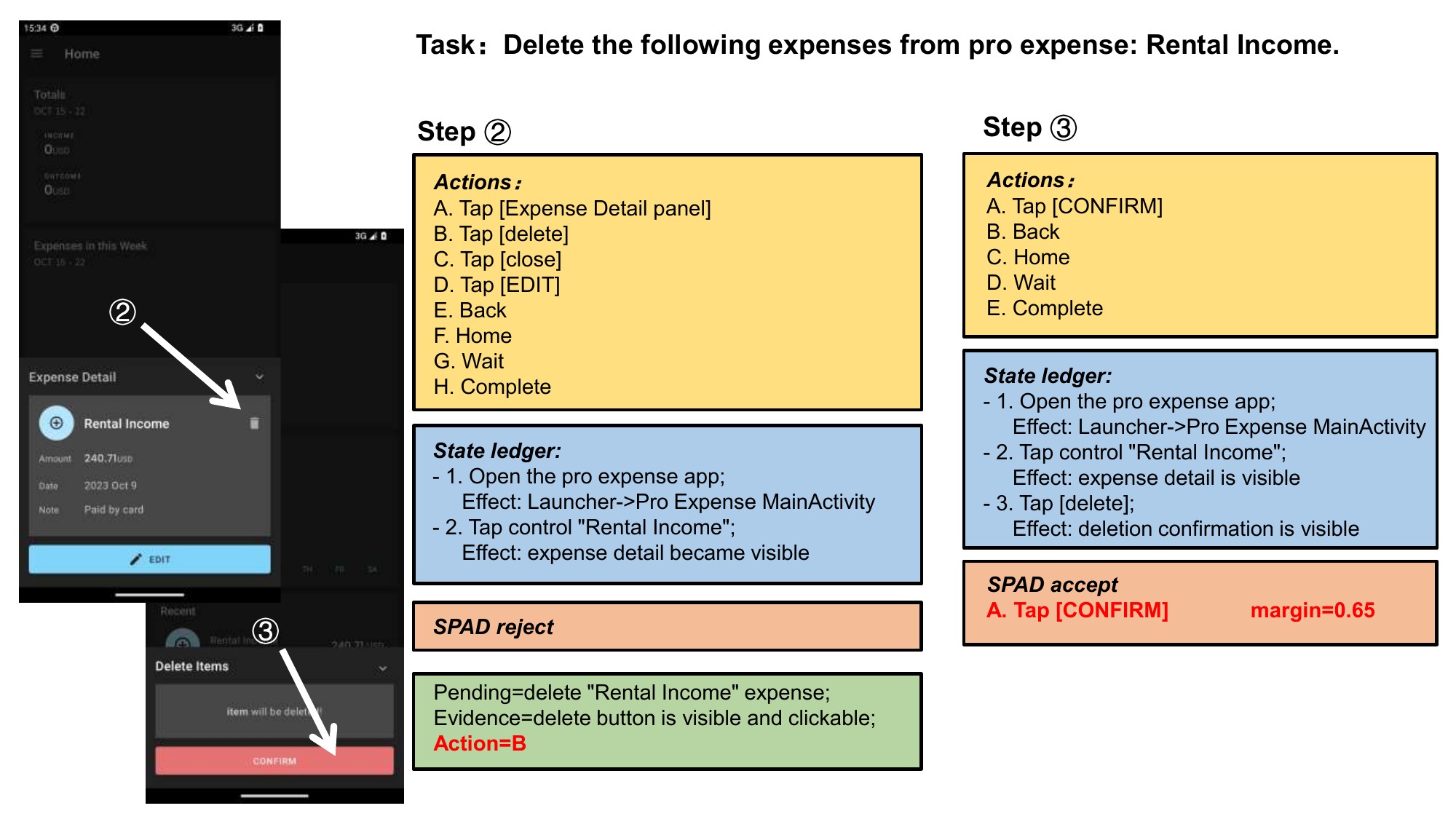}
    \caption{A representative DroidPrefill interaction showing the two
prefill-centric decision paths. At Step~\textcircled{2}, the first-glance
decision does not satisfy the SPAD condition, so DroidPrefill continues
with Structured Decision Refinement before
selecting the delete action. At Step~\textcircled{3}, the first-glance
decision passes the SPAD confidence test and is dispatched directly.
The candidate set and bounded state ledger provide the decision context
at both steps.}
    \label{fig:decision-example}
\end{figure}
After Semantic Candidate Compilation and Efficient Context Management,
DroidPrefill performs a single prefill pass over the bounded request.
Because the task, current UI, state ledger, and candidate semantics have
already been processed during prefill, the next action can be expressed
as a compact decision over candidate labels.

Figure~\ref{fig:decision-example} illustrates the two decision paths of
DroidPrefill. At Step~\textcircled{2}, the first-glance decision is
rejected by SPAD and the model proceeds through Structured Decision
Refinement before selecting the action. At Step~\textcircled{3}, the
first-glance decision satisfies the SPAD condition and is dispatched
directly from the prefill boundary.

DroidPrefill places the candidate list near the end of the request and
terminates at a dedicated decision boundary:
\begin{verbatim}
Candidate operations:
A. Tap [Internet; Networks available]
B. Press Back
C. Swipe down
D. Complete the task

Decision:
\end{verbatim}

After prefilling $x_t$, the runtime reads the next-token logit
associated with each legal label:
\begin{equation}
    z_{t,i}
    =
    f_{\theta}(x_t)[\ell_i],
    \qquad
    c_i\in\mathcal{C}_t,
    \label{eq:candidate-logit}
\end{equation}
where $\ell_i$ is the label assigned to candidate $c_i$.
The preferred candidate is
\begin{equation}
    \hat{c}_t
    =
    \arg\max_{c_i\in\mathcal{C}_t} z_{t,i}.
    \label{eq:first-glance-candidate}
\end{equation}
No candidate token is committed at this point; these logits provide an
early action preference directly from the prefilled context.
\paragraph{Self-Speculative Action Dispatch (SPAD)}
SPAD selectively dispatches decisions that are both confident and
structurally eligible.
Let $m_t$ denote the confidence margin used by SPAD, computed from the
candidate logits.
The dispatch condition is
\begin{equation}
    m_t \ge \tau
    \quad\land\quad
    \mathrm{Eligible}(\hat{c}_t,s_t,L_t)=\mathrm{TRUE}.
    \label{eq:selfspec-dispatch}
\end{equation}
When the condition holds, DroidPrefill executes $\hat{c}_t$ without
waiting for additional model decoding.

When the decision does not pass SPAD, DroidPrefill does not execute the
top-ranked candidate immediately.
Instead, it enters a short
\emph{Structured Decision Refinement} step.
The purpose of this continuation is to provide additional model
computation before committing to the final action, allowing the model
to reconsider the pending subgoal, inspect relevant evidence from the
current UI state, and resolve ambiguity among candidate actions.

The runtime blocks candidate labels at the first continuation position,
and continues
autoregressive decoding from the KV cache produced during the original
prefill.
Unlike open-ended chain-of-thought generation, Structured Decision
Refinement follows a compact format and is invoked only when the
prefill decision is not sufficiently confident. This conditional design is consistent with recent evidence that unconditional reasoning does not consistently improve mobile GUI-agent
performance~\cite{zhang2026cotgui}.
It therefore provides additional deliberation for difficult decisions
without making long-form generation a mandatory cost for every
interaction step.
The continuation follows the structured format:
\begin{verbatim}
Pending=<unfinished subgoal>;
Evidence=<evidence from the current state>;
Action=<candidate label>
\end{verbatim}

This structure constrains refinement along three dimensions.
First, refinement is conditional, so straightforward decisions do not
pay for additional autoregressive computation.
Second, the continuation uses a bounded output format rather than an
open-ended reasoning trace.
Third, the final decision remains grounded in the candidate set
constructed for the current UI state.
Structured Decision Refinement can therefore provide additional model
computation without reopening the action space or requiring the executor
to interpret arbitrary model-generated actions.

For an accepted SPAD decision, DroidPrefill executes the selected action
immediately while the same continuation may proceed in the background
for consistency checking and telemetry.
Thus, the current implementation reduces inference latency; it does
not rely on stopping the continuation to claim lower model compute.

\paragraph{Select-then-Fill.}
Most GUI operations are fully specified once their operation and target
have been selected.
Actions such as \textsc{Tap}, \textsc{Back}, \textsc{Home},
\textsc{Swipe}, \textsc{Wait}, and \textsc{Complete} therefore require
no additional free-form generation.
Only parameterized operations such as \textsc{Type} and \textsc{Speak}
require a textual value.

DroidPrefill separates operation selection from value generation. Let
$\sigma(\hat{c}_t)$ denote the optional value slot of the selected candidate:
\begin{equation}
  \begin{aligned}
    a_t &=
    \begin{cases}
        \Phi(\hat{c}_t),
        & \sigma(\hat{c}_t)=\varnothing,\\[1mm]
        \Phi(\hat{c}_t,v_t),
        & \sigma(\hat{c}_t)\ne\varnothing,
    \end{cases}
    \\
    v_t &=
    f_{\theta}^{\mathrm{fill}}
    (q,s_t,L_t,\hat{c}_t),
  \end{aligned}
  \label{eq:select-then-fill}
\end{equation}
where $\Phi$ binds the selected candidate and optional value to an
Android-executable action.
The fill stage uses the same model but cannot change the selected
operation or target.

Before execution, DroidPrefill verifies that the selected candidate
remains valid and that any required value is available.
After executing $a_t$, the system observes the next UI state and
deterministically updates the bounded state ledger before beginning the
next interaction step.

\begin{table*}[t]
\centering
\caption{Full AndroidWorld results.
Task success rate is reported separately for Easy tasks and the combined
Medium+Hard subset, while efficiency is micro-averaged per interaction
step over all model calls. QAIRT uses W4A16 and llama.cpp uses Q4\_0 on
the same smartphone.}
\label{tab:agent-performance}

\scriptsize
\setlength{\tabcolsep}{2.9pt}
\renewcommand{\arraystretch}{1.08}

\begin{tabular}{@{}lllllccrrr@{}}
\toprule
\multirow{2}{*}{\textbf{Observation}}
& \multirow{2}{*}{\textbf{Method}}
& \multirow{2}{*}{\textbf{Base Model}}
& \multirow{2}{*}{\textbf{Runtime}}
& \multirow{2}{*}{\textbf{Quantization}}
& \multicolumn{2}{c}{\textbf{Task SR} $\uparrow$}
& \multicolumn{3}{c}{\textbf{Efficiency over all tasks} $\downarrow$} \\
\cmidrule(lr){6-7}
\cmidrule(lr){8-10}

& & & &
& \shortstack{\textbf{Easy}}
& \shortstack{\textbf{Medium + Hard}}
& \shortstack{\textbf{Input tokens}}
& \shortstack{\textbf{Output tokens}}
& \shortstack{\textbf{Inference latency (s)}} \\
\midrule

\multirow{6}{*}{\shortstack[l]{A11y\\only}}
& T3A+
& Qwen3-4B-Instruct
& QAIRT
& W4A16
& 10.7 & 0.0
& 4023.0 & 318.6 & 24.2 \\

& V-Droid-offline
& Llama-3.1-8B\textsuperscript{$\dagger$}
& QAIRT
& W4A16
& 11.5 & 0.0
& 31583.8\textsuperscript{$\dagger$} & 47.1 & 10.6 \\

& \textbf{DroidPrefill-A11y}
& Qwen3-4B-Instruct
& QAIRT
& W4A16
& \textbf{23.0} & \textbf{3.6}
& \textbf{1830.4} & \textbf{26.7} & \textbf{4.6} \\

\cmidrule(lr){2-10}

& T3A+
& Qwen3-4B-Instruct
& llama.cpp
& Q4\_0
& 13.1 & 5.5
& 4917.4 & 249.4 & 34.8 \\

& V-Droid-offline
& Llama-3.1-8B\textsuperscript{$\dagger$}
& llama.cpp
& Q4\_0
& 26.2 & 3.6
& 41956.9\textsuperscript{$\dagger$} & 45.4 & 13.3 \\

& \textbf{DroidPrefill-A11y}
& Qwen3-4B-Instruct
& llama.cpp
& Q4\_0
& \textbf{37.7} & \textbf{7.3}
& \textbf{1833.4} & \textbf{33.0} & \textbf{5.2} \\

\midrule

\multirow{3}{*}{\shortstack[l]{Screenshot\\only}}
& Screenshot-only
& Qwen3-VL-4B-Instruct
& llama.cpp
& Q4\_0
& 14.8 & \textbf{7.3}
& 1315.8 & 46.3 & 16.1 \\

& Screenshot-only
& MAI-UI-2B
& llama.cpp
& Q4\_0
& 8.2 & 0.0
& 1052.5 & \textbf{22.1} & \textbf{13.3} \\

& Screenshot-only
& GUI-Owl-1.5-4B
& llama.cpp
& Q4\_0
& \textbf{16.3} & 3.6
& \textbf{897.5} & 46.9 & 18.6 \\

\midrule

\multirow{4}{*}{\shortstack[l]{A11y +\\screenshot}}
& M3A+
& Qwen3-VL-4B-Instruct
& QAIRT
& W4A16
& 14.8 & 0.0
& 4118.4 & 111.7 & 19.7 \\

& \textbf{DroidPrefill-Hybrid}
& Qwen3-VL-4B-Instruct
& QAIRT
& W4A16
& \textbf{23.0} & \textbf{1.8}
& \textbf{2313.1} & \textbf{32.3} & \textbf{15.0} \\

\cmidrule(lr){2-10}

& M3A+
& Qwen3-VL-4B-Instruct
& llama.cpp
& Q4\_0
& 24.6 & 3.6
& 5248.2 & 106.9 & 65.9 \\

& \textbf{DroidPrefill-Hybrid}
& Qwen3-VL-4B-Instruct
& llama.cpp
& Q4\_0
& \textbf{31.1} & \textbf{5.5}
& \textbf{2310.3} & \textbf{39.3} & \textbf{37.5} \\

\bottomrule
\end{tabular}

\vspace{2pt}
\parbox{\textwidth}{\scriptsize
\textsuperscript{$\dagger$} For V-Droid-offline, Llama-3.1-8B is used
only as the verifier; working-memory updates and action completion are
handled by a local Qwen3-4B-Instruct model on the same device.
Because V-Droid reuses a large cached prefix across calls, only
3,370.3 tokens per interaction step are actually prefetched on average.}

\end{table*}

\section{Implementation and Evaluation}
\label{sec:impl-eval}
\label{sec:implementation}
\label{sec:evaluation}

\subsection{Experimental Setup}
\label{sec:experimental-setup}

\subsubsection{Hardware and runtime.}

We use a split benchmark--agent setup for reproducible AndroidWorld
evaluation.
The Android emulator and benchmark controller run on the host PC to
support controlled task initialization, reliable state reset, action
injection, and success checking.
At each interaction step, the host transfers the raw UI observation
produced by the emulator to the smartphone and receives the executable
action selected by the agent.

The complete DroidPrefill agent runtime executes on a physical Xiaomi
17 smartphone powered by a Qualcomm Snapdragon 8 Elite Gen 5 SoC.
The smartphone performs UI parsing, Semantic Candidate Compilation,
Efficient Context Management, state-ledger maintenance, model
inference, action selection, value filling, and action validation.
For model inference, we evaluate two runtime paths provided by GenieX on the
smartphone NPU. The first uses the GenieX QAIRT runtime (QAIRT) with
precompiled W4A16 bundles through the QNN HTP backend. The second uses the
GenieX llama.cpp runtime (llama.cpp) with Q4\_0 GGUF models configured for
the Hexagon NPU. 
No cloud model, remote inference API, or host-side planning component
participates in the per-step decision process.

\subsubsection{Model deployment.}

For text-based agent configurations, we use
Qwen3-4B-Instruct-2507~\cite{qwen3}.
For configurations requiring visual input, we use
Qwen3-VL-4B-Instruct~\cite{qwen3vl2025}.
For brevity, we refer to the former as
Qwen3-4B-Instruct in the following tables.

The QAIRT path uses the released W4A16 model through Qualcomm
GENIE and the QAIRT/QNN HTP backend
~\cite{qualcomm-genie,qualcomm-qairt}.
The llama.cpp path uses a Q4\_0 model converted from the same released
checkpoint.
Unless otherwise stated, the main AndroidWorld evaluation uses the
released checkpoints without task-specific fine-tuning.
This setting isolates the effect of the agent architecture and inference
pipeline and keeps the main DroidPrefill system training-free.

\subsubsection{Benchmarks and metrics.}

We evaluate task effectiveness on the full AndroidWorld benchmark and
use its official \emph{Easy}, \emph{Medium}, and \emph{Hard} task
labels.
All systems use the same benchmark tasks, initial emulator states,
action executor, timeout policy, state-reset procedure, and success
checker.

Task effectiveness is measured by task success rate.
Efficiency metrics are micro-averaged over GUI interaction steps.
Across a task, we aggregate the input and output tokens from all model
calls and divide them by the number of GUI interaction steps.
This accounting includes auxiliary model calls such as summarization,
working-memory updates, and value filling when required by an agent.

We report \emph{inference latency}, measured along the model
critical path from the beginning of model-side processing until an
executable action becomes available.
For an SPAD-accepted decision, model computation that continues in the
background after action dispatch is excluded because it does not block
action execution.
The separate SPAD evaluation in
Section~\ref{sec:selfspec-analysis} analyzes this behavior in detail.

\subsubsection{Baselines.}

We restrict our comparison to representative mobile GUI-agent systems
whose model-inference pipelines can be executed locally on a smartphone.
Methods that fundamentally require proprietary cloud models, remote
multimodal APIs, or server-side planning are outside the offline
deployment setting targeted by DroidPrefill.

Within this scope, we compare against four representative agent
paradigms.

\begin{itemize}
    \item \textbf{T3A+}.
    T3A~\cite{androidworld2025} is a text-based generative GUI agent that consumes the task,
    structured UI representation, and interaction history and
    autoregressively generates the next action.
    To make its long requests feasible for mobile-NPU deployment, we apply
    the AndroidLab-style XML compressor~\cite{androidlab2025}; we refer to
    this adapted on-device version as T3A+.

    \item \textbf{M3A+.}
    M3A~\cite{androidworld2025} is a multimodal generative GUI agent that predicts the next action
    from the current screenshot, optionally together with structured UI
    information.
    We similarly apply the AndroidLab-style XML compressor
    ~\cite{androidlab2025} for mobile deployment and refer to the resulting
    adapted version as M3A+.
    
    \item \textbf{Screenshot-only.}
    Recent GUI-agent models can infer actions directly from screen images
    without requiring accessibility-tree or XML observations.
    We evaluate this paradigm using Qwen3-VL-4B-Instruct,
    MAI-UI-2B~\cite{maiui2025}, and
    GUI-Owl-1.5-4B~\cite{mobileagentv35_2026}.
    For each model, we use publicly available GGUF weights and run inference
    locally on the smartphone.
    To satisfy the input constraints of the mobile NPU runtime, screenshots
    are resized to the supported image resolution before inference.
    
    \item \textbf{V-Droid-offline.}
    V-Droid\cite{vdroid2026} constructs a discrete action space and uses a
    verifier to select the next operation.
    The original system relies on remote LLM calls for working-memory
    updates and generative action completion.
    In our tests, the released V-Droid checkpoint is specialized for the
    verification task and does not reliably follow instructions for these
    auxiliary generation tasks.
    We therefore retain the released V-Droid model as the verifier, while
    replacing the remote working-memory and action-completion calls with a
    locally deployed Qwen3-4B-Instruct model on the same smartphone.
    All verifier, working-memory, and action-completion model calls are
    included in the reported efficiency metrics.
\end{itemize}

\subsection{End-to-End GUI-Agent Performance}
\label{sec:eval-task-performance}

Table~\ref{tab:agent-performance} compares DroidPrefill with the locally
deployable baselines under different observation modalities, inference
runtimes, and quantization formats.

We evaluate three observation modalities that represent different
approaches to GUI perception and grounding.
\emph{A11y-only} agents operate on accessibility (A11y)-based
structured UI information, including element semantics, states,
hierarchical relationships, and grounding information, without using
screen images.
\emph{Screenshot-only} agents instead receive the current screenshot
and directly infer the next GUI action from visual input; spatial
operations such as tapping and swiping are grounded by predicting
screen coordinates rather than selecting A11y elements.
\emph{A11y+screenshot} combines the two sources in an
M3A multimodal interface.
A11y-derived element locations are mapped onto the screenshot using
Set-of-Mark annotations, which associate visual regions with
structured UI elements.
The model can therefore use the screenshot to perceive visual
appearance and layout while using the A11y representation to obtain
explicit element semantics, states, and grounding identities.

Within each controlled comparison, the systems use the same benchmark
environment, action executor, timeout policy, and success checker.
In particular, T3A+ and DroidPrefill use the same
Qwen3-4B-Instruct-2507 checkpoint within each runtime, providing a
controlled comparison of the agent inference pipeline.

\textbf{Task effectiveness.}
Under the A11y-only QAIRT/W4A16 setting, DroidPrefill improves Easy-task
success rate from 10.7\% with T3A+ to 23.0\%.
Under llama.cpp/Q4\_0, the corresponding Easy-task success rate improves
from 13.1\% to 37.7\%. The latter corresponds to a 24.6\% improvement,
the largest gain among the evaluated deployments.
These controlled comparisons indicate that replacing open-ended action
generation with grounded candidate selection and a bounded decision
interface can improve, rather than sacrifice, task effectiveness for
tasks within the capability of the deployed model.

\textbf{Difficulty breakdown.}
The improvement is concentrated on Easy tasks, while the gains on
Medium and Hard tasks are limited.
This behavior suggests that DroidPrefill primarily removes
inference- and interface-related failure modes for tasks that the
underlying on-device model is capable of solving.
For more difficult tasks, performance remains limited by the semantic
reasoning and long-horizon planning capability of the compact model.
DroidPrefill changes how the model is used on mobile hardware, but does
not remove the underlying model-capability ceiling.

\textbf{Inference efficiency.}
Under the A11y-only QAIRT/W4A16 setting, DroidPrefill reduces the
average input length from 4,023.0 to 1,830.4 tokens per interaction step
and reduces generated output from 318.6 to 26.7 tokens per step,
corresponding to reductions of 54.5\% and 91.6\%, respectively.
The resulting inference latency decreases from 24.2~s to
4.6~s per step, an 80.8\% reduction and a 5.2$\times$ speedup.
Under llama.cpp/Q4\_0, the inference latency further decreases
from 34.8~s to 5.2~s per step, corresponding to an 85.0\% reduction.

The large reduction in generated tokens is especially important on the
mobile NPU because, as shown in
Section~\ref{sec:prefill-decode}, autoregressive decode is substantially
more expensive than prompt prefill.
The end-to-end result therefore confirms the central design goal of
DroidPrefill: shifting the common-case GUI decision away from long
autoregressive generation substantially shortens the action critical
path.

\textbf{Runtime and quantization effects.}
The same agent can exhibit noticeably different task effectiveness
across the QAIRT/W4A16 and llama.cpp/Q4\_0 deployment paths.
For example, DroidPrefill-A11y achieves 23.0\% Easy-task success under
QAIRT/W4A16 and 37.7\% under llama.cpp/Q4\_0.
One possible contributor is activation quantization. Prior mobile-LLM
measurements show that W4A16 can improve NPU inference efficiency while
introducing model-dependent quality degradation compared with
weight-only quantization~\cite{npuready2026,mobilequant2024}.
However, because the comparison also changes the inference runtime and
quantization format, the current results do not isolate activation
quantization as the sole cause.
We therefore treat this difference as an observed system-level
runtime/quantization effect rather than making a causal attribution.

\textbf{Observation modality.}
Under our setting, adding visual observations does not
consistently improve task success and introduces substantial additional
latency.
A key reason is that visual input must operate under a much tighter
resolution--latency tradeoff than in server-side GUI agents.
AndroidWorld provides 1080$\times$2400 screenshots, but directly
processing such high-resolution images is prohibitively expensive on
our mobile deployment.
In our deployment diagnostics, full-resolution or near-full-resolution
inputs require more than 80~s for visual encoding, whereas resizing the
input to 512$\times$512 reduces visual processing to only a few seconds.

We therefore use a fixed 512$\times$512 visual input in the main
evaluation.
This operating point is also consistent with the statically compiled
visual profile supported by our NPU deployment, which limits the visual
token budget and input shape.
However, aggressive downsampling removes fine-grained spatial details
that are important for GUI grounding.
Our diagnostic runs illustrate this tradeoff: further reducing the
resolution to 224$\times$512 frequently produces malformed coordinate
outputs, while increasing it to 448$\times$1024 substantially increases
latency without reliably recovering correct grounding.
Thus, the visual models in our evaluation operate under a constrained
perception regime in which preserving high-resolution GUI details is
in tension with practical on-device inference latency.

Moreover, combining screenshots with accessibility information does not
consistently outperform either modality alone.
Although visual input provides appearance and layout information, it
also incurs vision-encoder overhead and additional visual tokens.
For compact on-device models, these costs can outweigh the benefit of
the additional modality, especially when the structured accessibility
representation already exposes much of the information required for
action selection.
These results suggest that the gains from richer multimodal input
observed with larger server-side GUI agents do not necessarily transfer
directly to resource-constrained on-device deployments.

\subsection{Ablation Analysis}
\label{sec:ablation}

We perform a leave-one-design-choice-out ablation by removing or
replacing one mechanism at a time while keeping the remaining pipeline
unchanged. Table~\ref{tab:ablation} summarizes the results.
\begin{table}[t]
\centering
\caption{Ablation on the AndroidWorld Easy
split. One mechanism is removed or replaced at a time while the remaining
pipeline is kept unchanged. Input/output tokens and inference
latency are averaged per GUI interaction step.}
\label{tab:ablation}

\scriptsize
\setlength{\tabcolsep}{3.2pt}
\renewcommand{\arraystretch}{1.10}

\begin{tabular}{@{}lcccc@{}}
\toprule
\textbf{Variant}
& \shortstack{\textbf{SR}\\\textbf{(\%)}}
& \shortstack{\textbf{Input}\\\textbf{tok./step}}
& \shortstack{\textbf{Output}\\\textbf{tok./step}}
& \shortstack{\textbf{Inference}\\\textbf{latency (s/step)}} \\
\midrule

\textbf{DroidPrefill}
& \textbf{37.7}
& \textbf{1800.4}
& \textbf{22.2}
& \textbf{4.4} \\

w/o SPAD
& 37.7
& 1789.0
& 32.4
& 5.6 \\

\makecell[l]{w/o SDR}
& 21.3
& 1340.5
& 213.7
& 24.1 \\

\makecell[l]{w/o CGUC}
& 31.1
& 2659.7
& 30.2
& 6.4 \\

\makecell[l]{w/o BSL}
& 21.3
& 1771.1
& 30.4
& 5.1 \\
\bottomrule
\end{tabular}
\end{table}

\textbf{Self-Speculative Action Dispatch (SPAD).}
We disable SPAD and require every decision to proceed through
Structured Decision Refinement before an action can be executed.
This variant therefore removes early action dispatch while preserving
the same candidate construction, context representation, and refinement
protocol.
Disabling SPAD leaves task success unchanged at 37.7\%, but increases
inference latency from 4.4~s to 5.6~s per step.
The output length also increases from 22.2 to 32.4 tokens per step
because every decision must wait for refinement.
This result confirms that SPAD primarily shortens the action critical
path without sacrificing task effectiveness.

\textbf{Structured Decision Refinement (SDR).}
We replace Structured Decision Refinement with free-form
chain-of-thought generation.
The model still receives the same task, UI context, state information,
and candidate set, but uncertain decisions are resolved using
unconstrained CoT rather than the structured refinement format.
This replacement causes the largest efficiency degradation:
output generation increases from 22.2 to 213.7 tokens per step, and
inference latency rises from 4.4~s to 24.1~s.
Meanwhile, task success drops from 37.7\% to 21.3\%.
These results show that unrestricted reasoning incurs substantial
autoregressive cost on the mobile NPU without improving GUI decisions,
whereas Structured Decision Refinement provides additional deliberation
within a compact and bounded output format.

\textbf{Candidate-grounded UI Compression (CGUC).}
We replace DroidPrefill's candidate-grounded UI compression with an
AndroidLab-style XML compression scheme.
The resulting compressed UI representation is used with the remaining
DroidPrefill pipeline unchanged, allowing us to isolate the effect of
preserving UI evidence according to the current candidate set.
With generic XML compression, the average input length increases from
1,800.4 to 2,659.7 tokens per step and inference latency increases
from 4.4~s to 6.4~s, while task success decreases from 37.7\% to
31.1\%.
This result indicates that candidate-grounded compression preserves a
more compact and decision-relevant UI representation than generic XML
compression under the same context budget.

\textbf{Bounded State Ledger (BSL).}
We replace the bounded state ledger with a conventional action history
that records recent executed actions.
The remaining context-construction and action-selection components are
kept unchanged.
Although the resulting input length remains similar
(1,771.1 versus 1,800.4 tokens per step), task success drops from
37.7\% to 21.3\%.
This suggests that the benefit of the bounded state ledger does not
primarily come from reducing token count; rather, its transition- and
state-aware representation preserves information that is important for
tracking task progress and making subsequent decisions.
\subsection{Self-Speculative Action Dispatch Analysis}
\label{sec:selfspec-analysis}

\begin{figure}[t]
    \centering
    \includegraphics[width=\linewidth]
    {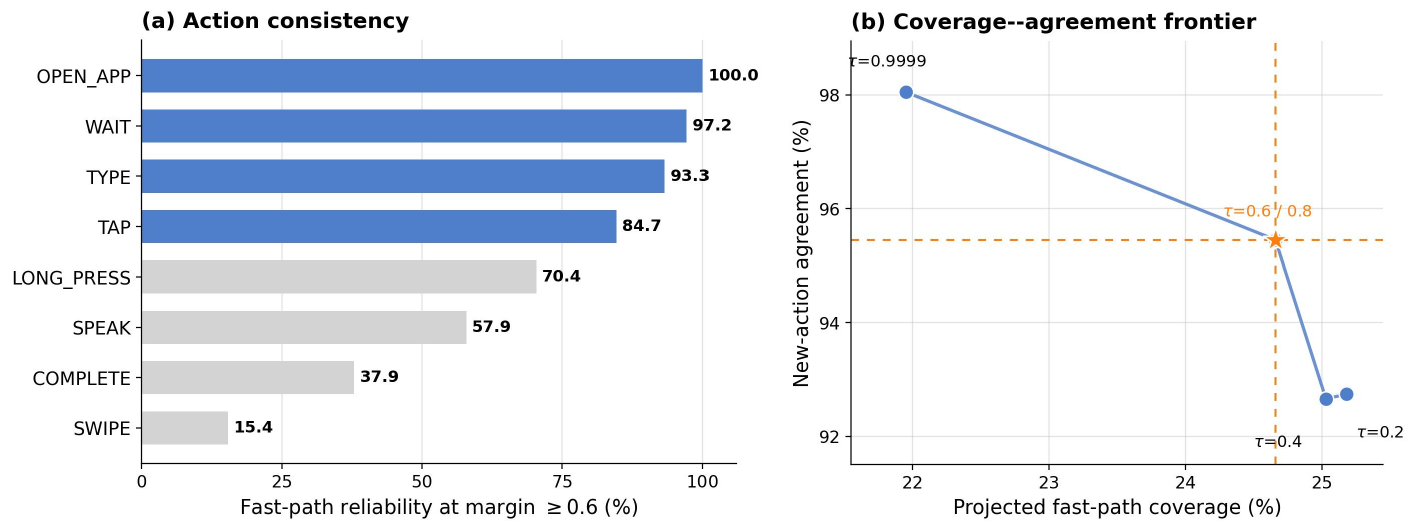}
    \caption{SPAD policy sensitivity on AndroidWorld split decisions.
    (a) First-glance agreement by operation type at $\tau=0.6$;
    highlighted operations are eligible for the fast path.
    (b) Fast-path coverage--agreement tradeoff as $\tau$ varies.
    Lower thresholds admit more decisions to the fast path but reduce
    agreement with the structured decision.
    The star marks the operating point used in our evaluation.}
    \label{fig:selfspec-policy}
\end{figure}

We evaluate how effectively SPAD can shorten the action critical path
by selectively dispatching reliable first-glance decisions.
SPAD uses an action-aware eligibility set together with a confidence
margin between the two highest-ranked candidate scores.
For an eligible action, we compute
\begin{equation}
    m_t = z_{t,(1)} - z_{t,(2)},
\end{equation}
where $z_{t,(1)}$ and $z_{t,(2)}$ denote the top two candidate scores.
The first-glance action is dispatched only when
\begin{equation}
    \mathrm{Eligible}(\hat{c}_t,s_t,L_t)=\mathrm{TRUE}
    \quad\land\quad
    m_t \ge \tau.
\end{equation}
We set $\tau=0.6$ in our evaluation.

Figure~\ref{fig:selfspec-policy} summarizes the resulting SPAD policy.
As shown in Figure~\ref{fig:selfspec-policy}(a), first-glance
reliability varies substantially across operation types.
\textsc{OpenApp}, \textsc{Wait}, \textsc{Type} and \textsc{Tap} achieve agreement
rates of 100.0\%, 97.2\%, 93.3\% and 84.7\%, respectively.
We therefore admit these operations to the fast path, while keeping
other operations ineligible.
For \textsc{Type}, SPAD dispatches only the operation and target
selection; textual value generation remains isolated in the subsequent
fill stage.

\textbf{Dispatch coverage and reliability.}
Figure~\ref{fig:selfspec-policy}(b) shows the coverage--agreement
tradeoff as the margin threshold varies.
At our selected operating point, $\tau=0.6$, the expanded policy
selects 24.7\% of decisions for the fast path.

Among the fast-path decisions covered by this policy, 95.5\% agree
with the original structured decision.
All disagreements occur in trajectories that already fail, while
all actions in the successful trajectories remain unchanged.
Under this trajectory-preserving replay, task success therefore remains
37.7\%.
No \textsc{Complete}, destructive, or answer-returning
operation enters the fast path.

\textbf{Latency benefit.}
Fast-path decisions become selection-ready after 0.9~s on average,
whereas decisions that proceed through Structured Decision Refinement
require 5.8~s.
Thus, for dispatched decisions, SPAD reduces the action-selection
critical path by 84.9\%.
Fast-path selection also emits no Structured Decision Refinement tokens.

Across all interaction steps, the replay projects the average output
generation to decrease from 32.4 to 22.2 tokens per step and the
inference latency to decrease from 5.6~s to 4.4~s per step,
corresponding to a 21.2\% overall latency reduction.
These results show that selectively exposing reliable operation classes
to the prefill-time fast path can substantially reduce decision latency
while preserving the successful trajectories observed in the original
execution.

\subsection{Real-Device End-to-End Validation}
\label{sec:real-device}

To verify that DroidPrefill operates as a fully on-device GUI agent
beyond the split AndroidWorld benchmark setup, we deploy the complete
pipeline on a physical smartphone and run representative tasks from
RealMobile~\cite{cao2026xiaomigui0}.
In this setting, UI observation, parsing, Semantic Candidate
Compilation, context construction, NPU inference, action selection,
value filling, and interaction execution all run on the same device
without a remote model or host-side planning component.

Across real-device tasks, DroidPrefill achieves an average end-to-end step latency of 5.0~s,
including both model inference and physical-device interaction overhead.
We use this experiment as an end-to-end feasibility validation rather
than a benchmark-scale accuracy evaluation.

\section{Limitations and Future Work}

DroidPrefill has several limitations.
First, fully on-device deployment currently relies on compact and
quantized models with bounded context, which limits semantic reasoning
and long-horizon planning on difficult GUI tasks.
Future work can use candidate- and SPAD-aware post-training to improve
both task accuracy and first-glance calibration.

Second, DroidPrefill primarily relies on accessibility-based UI
representations, which may be incomplete for visually rich or
custom-rendered interfaces.
Selective visual grounding could invoke the vision encoder only when
structured UI information is insufficient.

Third, SPAD currently depends on manually calibrated confidence
thresholds and action eligibility rules, while the bounded state ledger
may still lose information in very long tasks.
Learned risk-aware dispatch and hierarchical or retrieval-based memory
are promising extensions.

Finally, our evaluation focuses on Qualcomm smartphone NPUs and uses
representative rather than benchmark-scale real-device validation.
Broader evaluation across NPU platforms, applications, and sustained
energy and thermal conditions remains future work.

\section{Related Work}
\label{sec:related-work}

\subsection{Mobile GUI Agents}

Mobile GUI agents map user instructions and GUI observations to
executable actions.
AndroidWorld introduces representative text-based and multimodal
agents, including T3A and M3A, together with a dynamic benchmark for
mobile interaction~\cite{androidworld2025}.
AutoGLM, AppAgent, MobileGPT, and AutoDroid explore general-purpose
mobile automation, multimodal interaction, and memory-augmented task
execution~\cite{autoglm2024,appagent2025,mobilegpt2024,
autodroid2024}.

Recent work increasingly targets compact or on-device agents:
AutoDroid-V2 uses an on-device SLM with code generation, while
AgentCPM-GUI develops compact mobile-use agents with a concise action
space~\cite{autodroidv2_2025,agentcpmgui2025}.
UI-TARS develops native GUI agents for direct interaction with graphical
interfaces~\cite{uitars2025}, while Step-GUI further explores compact
GUI-agent models for mobile interaction~\cite{stepgui2025}.
V-Droid discretizes the GUI action space and uses a generative verifier
for action selection~\cite{vdroid2026}, whereas AgentProg addresses
growing interaction histories in long-horizon GUI tasks through
program-guided context management~\cite{agentprog2026}.
MobileExplorer studies fully on-device GUI-agent acceleration through
online UI exploration, and Parallel Speculative Execution explores
speculative execution for mobile GUI interaction
~\cite{mobileexplorer2026,pse2025}.

\subsection{On-Device LLM Inference}

Prior work improves on-device language-model inference through memory
management and heterogeneous execution.
LLM in a Flash reduces memory pressure through on-demand parameter
loading, while PowerInfer-2 jointly manages smartphone compute, memory,
and storage resources~\cite{llminaflash2024,powerinfer2_2024}.
LLM.NPU and ShadowNPU optimize LLM inference around mobile NPUs through
heterogeneous scheduling and NPU-centric system--algorithm co-design
~\cite{llmnpu2025,shadownpu2026}.
Recent measurement work further characterizes efficiency bottlenecks
across mobile LLM runtimes and hardware backends
~\cite{npuready2026}, while MobileQuant studies mobile-friendly
quantization and its efficiency--quality tradeoff
~\cite{mobilequant2024}.
Recent work also explores context management for on-device LLM
services~\cite{ondevicellmaas2026}.
Agent-X accelerates end-to-end on-device agent workloads through
prefix-cache-aware prompt restructuring and agent-specific speculative
decoding~\cite{agentx2026}.
More general speculative and adaptive inference techniques reduce the
cost of autoregressive generation
~\cite{leviathan2023speculative,elhoushi2024layerskip,
schuster2022calm}.

\section{Conclusion}
\label{sec:conclusion}

This paper revisits fully on-device mobile GUI agents from the
perspective of mobile-NPU execution.
We show that conventional GUI-agent pipelines expose two key
workload--hardware mismatches: substantial autoregressive decoding despite
the large prefill--decode throughput gap of mobile NPUs, and dynamically
varying contexts that conflict with statically provisioned context profiles.

We present \emph{DroidPrefill}, a prefill-centric architecture that
reshapes GUI-agent inference around these execution characteristics.
DroidPrefill combines \emph{Efficient Context Management} to preserve
decision-critical information within a bounded context budget with
\emph{Prefill-Centric Inference} to move common-case action selection
toward the prefill boundary.
SPAD dispatches confident decisions early, while uncertain decisions use
short Structured Decision Refinement and textual values are generated only
when required.

Our evaluation on commercial smartphones shows that DroidPrefill reduces
per-step inference latency by up to 85.0\% and improves task success rate
by up to 24.6\%.
We further validate the complete observation--decision--execution loop
on a physical smartphone using representative RealMobile tasks.
These results show that efficient fully on-device GUI agents require
redesigning the agent workload around mobile-hardware execution
characteristics, not merely model compression or faster runtimes.

\bibliographystyle{ACM-Reference-Format}
\bibliography{references}

\end{document}